\documentclass[11pt]{article}

\usepackage{graphics}

\makeatletter
  
  \@addtoreset{equation}{section}
\makeatother

\newcommand{\lam}{\lambda_0}
\newcommand{\lag}{2{\cal L(C)}}

\newcommand{\delxy}{\delta_{x_0,y_0}}
\newcommand{\delxpoy}{\delta_{x_{0}+1,y_0}}
\newcommand{\delxmoy}{\delta_{x_{0}-1,y_0}}
\newcommand{\delxpty}{\delta_{x_{0}+2,y_0}}
\newcommand{\delxmty}{\delta_{x_{0}-2,y_0}}
\newcommand{\delxz}{\delta_{x_{0},0}}
\newcommand{\delxo}{\delta_{x_{0},1}}
\newcommand{\delxTmo}{\delta_{x_{0},L-1}} 

\newcommand{\px}{(\mbox{\bf{p}},x_0)}
\newcommand{\pxpo}{(\mbox{\bf{p}},x_0+1)}
\newcommand{\pxmo}{(\mbox{\bf{p}},x_0-1)}
\newcommand{\pxpt}{(\mbox{\bf{p}},x_0+2)}
\newcommand{\pxmt}{(\mbox{\bf{p}},x_0-2)}
\newcommand{\py}{(\mbox{\bf{p}},y_0)}
\newcommand{\pypo}{(\mbox{\bf{p}},y_0+1)}
\newcommand{\pymo}{(\mbox{\bf{p}},y_0-1)}

\newcommand{\pxy}{(\mbox{\bf{p}};x_0,y_0)}
\newcommand{\pyx}{(\mbox{\bf{p}};y_0,x_0)}

\newcommand{\phix}{\phi_a(x_0)}
\newcommand{\phixpo}{\phi_a(x_0+1)}
\newcommand{\phiy}{\phi_a(y_0)}

\newcommand{\shouleft}{\left ( \mbox{\hspace{-1mm}} \frac{}{} \mbox{\hspace{-1mm}} \right.}
\newcommand{\shouright}{\left. \mbox{\hspace{-1mm}} \frac{}{} \mbox{\hspace{-1mm}} \right)}
\newcommand{\chuuleft}
           {\mbox{\hspace{-1mm}} \left \{ \mbox{\hspace{-1mm}} \frac{}{} \mbox{\hspace{-1mm}} \right.}
\newcommand{\chuuright}
           {\left. \mbox{\hspace{-1mm}} \frac{}{} \mbox{\hspace{-1mm}} \right\}}
\newcommand{\daileft}
           {\mbox{\hspace{-1mm}}\left [ \mbox{\hspace{-1mm}} \frac{}{} \mbox{\hspace{-1mm}} \right.}
\newcommand{\dairight}
           {\left. \mbox{\hspace{-1mm}} \frac{}{} \mbox{\hspace{-1mm}} \right]}

\title{
A perturbative determination of O($a$) boundary
improvement coefficients for the Schr\"odinger Functional coupling
at 1-loop with improved gauge actions
}

\author{
Shinji Takeda, Sinya Aoki and Kiyotomo Ide \\
Institute of Physics, University of Tsukuba \\
Ibaraki 305-8571, Japan
}

\begin{document}

\maketitle

 \begin{abstract}
We determine O($a$) boundary improvement
coefficients up to 1-loop level
for the Schr\"odinger Functional coupling
with improved gauge actions including
plaquette and rectangle loops.
These coefficients are required to implement
1-loop O($a$) improvement in 
full QCD simulations for the coupling with the improved gauge actions.
To this order, lattice artifacts of 
step scaling function for each improved gauge action
are also investigated.
In addition, passing through the SF scheme,
we estimate the ratio of $\Lambda$-parameters
between the improved gauge actions and the plaquette action
more accurately.

 \end{abstract}


 \section{Introduction}
  \label{sec:introduction}

The $\overline{\rm{MS}}$-scheme now becomes the standard renormalization 
scheme for the definition of the strong coupling constant.
The measured coupling constant $\alpha_{\rm{s}}$ in some experiment
at relatively high energy is converted to $\alpha_{\overline{\rm{MS}}}$ 
at some representative scale by perturbation theory. 
The current world average of such estimates gives
$\alpha_{\overline{\rm{MS}}}(m_{Z}=91.19 {\rm GeV}) \approx 0.11$.
Lattice QCD calculations, on the other hand, have a potential
ability to determine the the strong coupling constant from the
experimental inputs at low (hadronic) energy scale.
In order to compare the coupling constant obtained at low energy by the
lattice calculations with $\alpha_{\overline{\rm{MS}}}$  obtained
at high energy, 
the Schr\"odinger Functional scheme has been proposed by the ALPHA 
collaboration \cite{sforiginal}, 
and the scheme is shown to be successful.
At present, the results on the running coupling constant
of two massless flavor QCD are reported \cite{first,recent}.

In the real world there are three light quarks.
QCD simulations including three dynamical quark
effects are thus required to understand the low energy QCD dynamics.
Our ultimate goal is to estimate $\alpha_{\overline{\rm{MS}}}$
from $N_{\rm{f}}=3$ QCD simulations.
Recently, CP-PACS/JLQCD collaborations
have started the $N_{\rm{f}}=3$ QCD simulation employing
an exact fermion algorithm developed for odd number
of quark flavors \cite{ishikawa,okawa,cswnonperturbative}.
In particular, noteworthy results have been obtained in \cite{okawa}:
There exists strong lattice artifacts associated with
the phase transition in the $N_{\rm{f}}=3$ QCD simulation
with the combination of the plaquette gauge action and O($a$)
improved Wilson quark action, while
such lattice artifacts are absent for the
renormalization group (RG) improved gauge action.
Hence, the collaborations have decided to adopt
the combination of the RG improved gauge action and O($a$) improved Wilson 
quark action for the $N_{\rm{f}}=3$ QCD simulations 
to obtain $\alpha_{\overline{\rm{MS}}}$.

As a first step of our program,
we study a Schr\"odinger Functional coupling
with improved gauge actions in perturbation theory.
In particular
we perturbatively calculate the O($a$) boundary improvement coefficients 
at 1-loop level with improved gauge actions for the pure SU(3) gauge theory.
Combining this with the 1-loop results for the fermionic sector \cite{sint},
we determine the O($a$) boundary improvement coefficients at 1-loop,
which can be used for the dynamical quark simulations in the future.

The rest of this paper is organized as follows.
In Sect 2, after a brief reminder of the Schr\"odinger Functional scheme and 
its extension to improved gauge actions,
we specify an action used for latter calculations,
and discuss the O($a$) boundary counterterm.
In Sect 3 the Schr\"odinger Functional
coupling constant is defined and formula for 
a determination of the O($a$) boundary improvement 
coefficients are given.
The 1-loop computation is outlined in Sect 4, and
the results of the O($a$) boundary improvement coefficients
are summarized in Sect 5.
Our conclusion is given in the last section, 
together with a discussion on the 
lattice artifact of the step scaling function.

 \section{Preliminaries}
  \label{sec:setup}

  \subsection{Schr\"odinger Functional}
   \label{sec:SF}

It has been shown by ALPHA collaboration 
that the Schr\"odinger Functional (SF) scheme is a powerful tool to
probe energy evolutions of some physical quantities
and to compute improvement coefficients as well as renormalization constants.
In the SF scheme, the theory is defined on a finite box of 
size $L^3\times T$ with periodic boundary conditions in 
the spatial directions and Dirichlet boundary
conditions in the time direction.
In the pure SU($3$) gauge theory with Wilson plaquette action $S[U]$,
the partition function in the SF scheme (in the case that $T=L$) is given by
    \begin{equation}
      {\cal Z} 
      =
      \int D[U] e^{ - S[U] },
    \end{equation}
where link variables $U(\mu,x)$ for gauge fields satisfy boundary
conditions
    \begin{equation}
      \left.U(x,k)\right|_{x_0 = 0} 
      = 
      \exp \{ a C \} ,\mbox{\hspace{5mm}}
      \left.U(x,k)\right|_{x_0 = L} 
      = 
      \exp \{ a C^{\prime} \}.
     \label{eqn:CC}
    \end{equation}
Here $a$ is lattice spacing, and
$C$, $C^{\prime}$ are diagonal traceless  matrices, which depend on 
background field parameters $\eta$ and $\nu$ \cite{su3}.
It is shown\cite{sforiginal} that the minimum of $S[U]$ is given by
the lattice background field $U(x,\mu)=V(x,\mu)$, where
    \begin{equation}
      V(x,0) 
      = 
      1,
      \mbox{\hspace{3mm}}   
      V(x,k)
      = 
      V(x_0),
     \label{eqn:V}
    \end{equation}
with
    \begin{eqnarray}
      V(x_0) 
      &=& 
      \exp \{a b(x_0)\} ,\\
      b(x_0) &=& \frac{1}{L} [(L - x_0) C + x_0 C^{\prime}].
    \end{eqnarray}
This background field represents a 
constant electric field.

An extension of the SF scheme to improved gauge actions
was first considered by Klassen \cite{klassen}.
The transfer matrix construction \cite{imptransfer} was 
adopted in the discussion. In this formulation,
Each boundary consists of two time slices, to achieved the tree-level 
O($a^2$) improvement.

In this paper, however, we adopt the formulation proposed by 
Aoki, Frezzotti and Weisz \cite{aokiweisz},
where each boundary consists of only one time slice and
the tree-level O($a$) improvement is achieved:
Dynamical variables to be integrated over are independent on the form of the 
action, plaquette or improved ones, and are given by
the spatial link variables $U(k,x)$ with $x_0 = a,\cdots,L-a$
and temporal link variables $U(0,x)$ with $x_0 = 0,\cdots,L-a$
on a cylinder with volume $L^3 \times L$.
This formulation is implemented more easily in numerical simulations.

The background field in eq.(\ref{eqn:V}) gives the local minimum
of the theory in both cases \cite{klassen} and \cite{aokiweisz}.
It has not been theoretically proved, however, that eq.(\ref{eqn:V}) is the
absolute minimum for the improved gauge actions.
This is checked only numerically \cite{klassen}.


  \subsection{Gauge action}
   \label{sec:gaugeaction}

Our improved action includes plaquette and rectangle loops,
and is given by
    \begin{equation}
      S_{\rm{imp}}[U] 
      = 
      \frac{1}{g^2_0} 
      \sum_{{\cal C} \in {\cal S}_0} W_0({\cal C},g^2_0) \lag
    + \frac{1}{g^2_0} 
      \sum_{{\cal C} \in {\cal S}_1} W_1({\cal C},g^2_0) \lag,
     \label{eqn:impaction}
    \end{equation}
with
    \begin{equation}
      {\cal L(C)} 
      = 
      \mbox{ReTr}[I - U({\cal C})],
    \end{equation}
where $U({\cal C})$ is the ordered product of 
the link variables along loop ${\cal C}$
contained in the set ${\cal S}_0$(plaquette) or ${\cal S}_1$(rectangular).
${\cal S}_0$ and ${\cal S}_1$ consist of 
all loops of the given shape which can be 
drawn on the cylindrical lattice with volume $L^3 \times L$.
The loops involve only the ``dynamical links'' in the
sense specified above, and spatial links
on the boundaries at $x_0 = 0$ and $x_0 = L$.
In particular, rectangles protruding from the boundary
of the cylinder are not included.

One has to choose appropriate
boundary weights to achieve the 1-loop level O($a$)
improvement for observables involving derivative
with respect to the boundary.
Among various choices to achieve this, ours is given as follows.
    \begin{equation} 
      W_0({\cal C},g^2_0) 
      = 
      \left\{
       \begin{array}{ll}
         c_{\rm{s}}(g^2_0) 
       & \mbox{for } {\cal C} \in P_{\rm{s}} : 
          \mbox{Set of plaquettes that lie on one of} 
       \\
       &
          \mbox{\hspace{19mm} the boundaries, } 
       \\      
         c_0 c^P_{\rm{t}}(g^2_0) 
       & \mbox{for } {\cal C} \in P_{\rm{t}} : 
          \mbox{Set of plaquettes that just touch one} 
       \\
       &
          \mbox{\hspace{19mm} of the boundaries, } 
       \\      
         c_0        
       & \mbox{for } {\cal C} \in P_{\rm{other}} : 
          \mbox{otherwise, } 
       \end{array}
            \right.
     \label{eqn:W0}
    \end{equation} 
    \begin{equation} 
      W_1({\cal C},g^2_0) 
      = 
      \left\{
       \begin{array}{ll}
         0          
       & \mbox{for } {\cal C} \in R_{\rm{s}} : 
          \mbox{Set of rectangles that lie completely} 
       \\
       &
          \mbox{\hspace{19mm} on one of the boundaries, } 
       \\
         c_1 c^R_{\rm{t}}(g^2_0) 
       & \mbox{for } {\cal C} \in R^2_{\rm{t}} : 
          \mbox{Set of rectangles that have exactly 2} 
       \\
       &
          \mbox{\hspace{19mm} links on a boundary, } 
       \\     
         c_1        
       & \mbox{for } {\cal C} \in R_{\rm{other}} : 
          \mbox{otherwise, }     
       \end{array}
            \right.
     \label{eqn:W1}
    \end{equation} 
with 
    \begin{eqnarray}
      c_0 c^P_{\rm{t}}(g^2_0)
      & = & 
      c_0 ( 1 + c^{P(1)}_{\rm{t}} g^2_0 + O(g^4_0) ), \\
      c_1 c^R_{\rm{t}}(g^2_0)
      & = & 
      c_1 ( 3/2 + c^{R(1)}_{\rm{t}} g^2_0 + O(g^4_0) ),
   \end{eqnarray}
where coefficients $c_0$ and $c_1$ of the improved gauge action are 
normalized such that $c_0 + 8 c_1 = 1$.
We call $c^P_{\rm{t}}(g^2_0)$ and $c^R_{\rm{t}}(g^2_0)$
O($a$) boundary improvement coefficients.
So far, 1-loop coefficients 
$c^{P(1)}_{\rm{t}}$ and $c^{R(1)}_{\rm{t}}$ 
are independent each other.
Weight factors $W_i({\cal C},g^2_0)$, which include loop corrections,
becomes the Choice B of \cite{aokiweisz} in the weak coupling limit.
The Choice B achieves the tree-level O($a$) improvement
and, at the same time, the lattice background field $V$ in eq.(\ref{eqn:V})
satisfies the equation of motion obtained by the 
variation of dynamical links.

Incidentally, we discuss
the O($a$) boundary counterterm from
a different point of view.
Here, it is assumed that the plaquette loops 
and rectangle loops 
which lie completely on the cylinder $L^3 \times L$
are included in the
action, 
and that each boundary consists of one time slice only.
As explained in \cite{sforiginal},
at order $a$ in the pure gauge theory,
there are two possible boundary counterterms,
$a^4 \mbox{Tr} \{ F_{0k}F_{0k} \}$ and $a^4 \mbox{Tr} \{ F_{kl}F_{kl} \}$,
each of which are summed over the $x_0 = 0$ or $x_0 = L$ hyper plane.
Since the latter boundary term vanishes in the case of Abelian constant 
boundary field, in the following, we consider only the former.
In this case,
we have three candidates for O($a$) boundary
counterterms which respect lattice symmetries,
    \begin{enumerate}
      \item spatial sum of time-like plaquette loop
            that just touches one of the boundaries,
      \item spatial sum of rectangle loop that
            has exactly 2 links on a boundary,
      \item spatial sum of rectangle loop that
            has exactly 1 link on a boundary,
    \end{enumerate} 
to satisfy one condition, the O($a$) improvement condition.
Therefore, for simplicity, we can take a trivial weight for the term 3, and
we still have one degree of freedom for the choice of the boundary terms.
At the tree level, however, the background field given in eq.(\ref{eqn:V})
must satisfy the equation of motion\footnote{ The equation of motion 
for the plaquette action is trivially satisfied. }, 
so that one has to take $ c_t^P =1$ and $c_t^R=3/2$ (Choice B).
Since no such an extra constraint exists for the 1-loop boundary terms,
we can freely set a relation between
$c^{P(1)}_{\rm{t}}$ and $c^{R(1)}_{\rm{t}}$,
which will be given in the next section.


 \section{SF coupling and  O($a$) boundary improvement coefficients}
  \label{sec:ct}
 
The SF with the improved gauge action is given by
    \begin{equation}
      {\cal Z}
      =
      e^{- \Gamma}
      =
      \int D[U] e^{ - S[U] },
     \label{eqn:SFimp}
    \end{equation}
where $S[U]=S_{\rm imp}[U]$.
We require the same boundary condition, eq.(\ref{eqn:CC}), for       
the link variables as in the case of the Wilson plaquette action.
In perturbative calculations,
there are two main concerns to note :
one is whether the background field given by eq.~(\ref{eqn:V}) 
corresponds to the absolute  minimum of the  action,
and the other is the gauge fixing.
For the latter, we used the covariant gauge
fixing procedure outlined in \cite{sforiginal}.
The former statement is positively proved in \cite{sforiginal}
for Wilson plaquette action.
Unfortunately the statement has not been proved yet in the case of
the improved gauge actions, since the proof in \cite{sforiginal} is not 
applicable to these cases.
In \cite{klassen}, however, it has been numerically checked 
that the background field given in eq.~(\ref{eqn:V}) corresponds to the minimum
for a large class of improved actions, hence we assume 
this in our perturbative calculations.

In a neighborhood of the background field $V$, any link variables $U$
can be parametrized by
    \begin{equation}
      U(x,\mu)
      =
      \exp \{ g_0 a q_{\mu}(x) \} V(x,\mu),
    \end{equation}
where $q_{\mu}$ are quantum fields.
The SF coupling is defined through
the free energy $\Gamma$ in eq.~(\ref{eqn:SFimp})
    \begin{equation}
      \bar{g}^2_{\rm{SF}}(L)
      =
      \left.
      \frac{\Gamma^{\prime}_0}{\Gamma^{\prime}} 
      \right|_{\eta=\nu=0},
    \end{equation} 
where $\Gamma^{\prime}$ is the derivative with
respect to $\eta$. $\Gamma^{\prime}_0$
is a normalization constant
    \begin{eqnarray}   
      \Gamma^{\prime}_0 
      & = & 
      \left.
      \frac{\partial}{\partial \eta}
        g^2_0 S[V]
      \right|_{g^2_0=0}
      \nonumber \\ 
      & = & 
      \frac{\partial}{\partial \eta}
      \left.
        \left[ 
          c_0 
           \left\{ 
             \sum_{P_{\rm{t}}} \lag 
           + \sum_{P_{\rm{other}}} \lag 
           \right\}
        + c_1 
           \left\{ 
             \frac{3}{2} \sum_{R^2_{\rm{t}}} \lag 
           + \sum_{R_{\rm{other}}} \lag 
           \right\} 
        \right] 
      \right|_{U=V}
      \nonumber \\ 
      & = & 
      12 \left(
           \frac{L}{a}
         \right)^2 
         \left[
           c_0 
           \shouleft 
             \sin 2 \gamma
           + \sin   \gamma
           \shouright
           + 4 c_1 
           \shouleft 
             \sin 4 \gamma
           + \sin 2 \gamma
           \shouright              
         \right],  
    \end{eqnarray}
where $\gamma$ is given in appendix \ref{sec:inversepropagator}.

Let us discuss the perturbative expansion of the SF coupling. 
If we require that 
$c^{P(1)}_{\rm{t}}$ and $c^{R(1)}_{\rm{t}}$ 
satisfy (this is possible by using the last degree of freedom
as mentioned in the end of the previous section)
    \begin{equation}
      c^{R(1)}_{\rm{t}}
      =
      2 c^{P(1)}_{\rm{t}},
     \label{eqn:relation}
    \end{equation}
and introduce $c_{\rm{t}}^{(1)}$
    \begin{equation}
      c_{\rm{t}}^{(1)}
      =
      c_0 c^{P(1)}_{\rm{t}}
    + 4 c_1 c^{R(1)}_{\rm{t}}
      =
      c_{\rm{t}}^{P(1)},
     \label{eqn:ct}
    \end{equation}
then one obtains the relation between $S[V]$ and $\Gamma_0$
    \begin{equation}
      S[V]
      =
      \left( 
      \frac{1}{g^2_0} + \frac{2a}{L} c^{(1)}_{\rm{t}} 
      \right)  \Gamma_0 
       + O(g^2_0).
     \label{eqn:Sgamma} 
    \end{equation}
Using the eq.(\ref{eqn:Sgamma}),
the perturbative expansion of $\bar{g}^2_{\rm{SF}}(L)$
is given by
    \begin{eqnarray}
      \bar{g}^2_{\rm{SF}}(L)
      =
      g^2_0 + m^{(1)}_1(L/a) g^4_0 + O(g^6_0), 
    \end{eqnarray} 
with 
    \begin{equation}
      m^{(1)}_1(L/a)
      =
    - \frac{2a}{L} c^{(1)}_{\rm{t}}
    + m^{(0)}_1(L/a),
     \label{eqn:ctm1}
    \end{equation}
where $m^{(0)}_1(L/a)$
is the 1-loop correction to the SF coupling, calculated
with the tree-level O($a$) boundary coefficients,
and the detail of the calculation
will be given in the next section.
The value of the 1-loop coefficient $c^{(1)}_{\rm{t}}$
is determined by the condition that 
the dominant part of the scaling violation of
$m^{(1)}_1(L/a)$ should be proportional to $(a/L)^2$,
and then 
$c^{P(1)}_{\rm{t}}$ and $c^{R(1)}_{\rm{t}}$ 
are uniquely given by eq.(\ref{eqn:relation}) and eq.(\ref{eqn:ct}).

 \section{Calculation of the 1-loop coefficient}
  \label{sec:m1}

In the following, we choose lattice unit (i.e. $a=1$).
According to the unpublished note \cite{weisz},
we have used $I^{a}$ $(a = 1,2,\cdots,8)$ as a basis of Lie algebra of SU(3)
in the presence of the background field.
Their explicit form can be found in \cite{kurth}.
Decomposing in a basis $I^a$
     \begin{equation}
       q_{\mu}(x)
       =  
       \sum_{a}
       \tilde{q}^{a}_{\mu}(x) I^a,
     \end{equation}
the quantum fields $q_{\mu}$ are Fourier transformed with respect to spatial 
momenta as
     \begin{eqnarray}
       \tilde{q}^a_{0}(x)
       & = & 
       \frac{1}{L^3} \sum_{\mbox{\scriptsize{\bf{p}}}}
             e^{i {\bf{px}}}
             \tilde{q}^{a}_{0}({\bf{p}},x_0),
       \\
       &   & 
       \nonumber \\
       \tilde{q}^a_{k}(x)
       & = & 
       \frac{1}{L^3} \sum_{\mbox{\scriptsize{\bf{p}}}}
             e^{i {\bf{px}}}
             e^{\frac{i}{2}( p_k + \phix )}
             \tilde{q}^{a}_{k}({\bf{p}},x_0),
     \end{eqnarray}                   
where the phase $\phix$ is given in appendix \ref{sec:inversepropagator}.
In terms of $\tilde{q}^a_{\mu}(x)$,
the quadratic part of the improved gauge action 
eq.(\ref{eqn:impaction}) takes the form
    \begin{equation}
      S^{(0)}_{\rm{imp}} 
      = 
      \frac{1}{L^3} 
      \sum_{{\mbox{{\scriptsize{{{\bf{p}}}}}}}}
      \sum^{T-1}_{x_0,y_0=0}
      \sum_a
      \tilde{q}^{\bar{a}}_{\mu}(-{\bf{p}},x_0)
      K^{a}_{\mu \nu}({\bf{p}};x_0,y_0)
      \tilde{q}^{a}_{\nu}({\bf{p}},y_0),
     \end{equation} 
with the condition
     \begin{equation}
       \left. 
       q_k({\bf{p}},x_0) 
       \right|_{x_0 = 0}
       = 
       0, 
       \mbox{\hspace{2mm} for $k = 1,2,3 $}.
     \end{equation}
The explicit form of the inverse propagator $K^a_{\mu \nu}$ 
is given in~\cite{aokiweisz}, and also in appendix \ref{sec:inversepropagator}.

The 1-loop correction $m^{(0)}_1(L)$
is given by
     \begin{equation}
      m^{(0)}_1(L)
      =
      \left.
      - \frac{1}{\Gamma^{\prime}_0} 
      \frac{\partial}{\partial \eta}
      \left[ \frac{1}{2} \ln \mbox{Det} K 
                       - \ln \mbox{Det} \Delta_0 \right]
      \right|_{\eta=\nu=0},
     \label{eqn:impm1}
    \end{equation}   
where the determinant for the quantum field sector (the first term in the 
right hand side of the equation) is taken with respect to
the spatial momentum {\bf{p}}, the time $x_0$, the Lie algebra sector $a$ and 
Lorentz index $\mu$.
The second term in the right hand of eq.(\ref{eqn:impm1})
represents a contribution from the ghost sector.
Here we will exclusively consider the quantum field sector,
since the contribution from the ghost sector to the 1-loop correction
is same as in the case of Wilson plaquette action.
Our boundary condition for the temporal component $q_0$
is different from that in \cite{sforiginal},
so that the
``non-uniform'' contribution in the gauge fixing term
remains in the inverse propagator $K$ 
(see appendix \ref{sec:inversepropagator}).

   \begin{table} [p] 
\vspace{-1cm}
    \begin{center} 
     \begin{tabular}{|c|c|c|c|} 
       \hline \hline
       $L$ & Iwasaki action & LW action & DBW2 action 
       \\ \hline
6 &  0.0865021015584032  & 0.3843092560841445  & -0.2542597063902088 \\
7 &  0.1026697312426737  & 0.4061279685078025  & -0.2517151449619943 \\ 
8 &  0.1171638577366678  & 0.4249279311929165  & -0.2462340808659547 \\
9 &  0.1303628849788211  & 0.4414718008748639  & -0.2394316692834335 \\
10&  0.1424404981803593  & 0.4562496675217995  & -0.2322864496797361 \\
11&  0.1535565273026473  & 0.4696045216347286  & -0.2251762317750280 \\
12&  0.1638489264935022  & 0.4817875167578513  & -0.2182314272034601 \\
13&  0.1734301653189940  & 0.4929883879248706  & -0.2114993870451222 \\
14&  0.1823916642888328  & 0.5033540747374835  & -0.2049961094379959 \\
15&  0.1908085280119190  & 0.5130007428050310  & -0.1987234947986162 \\
16&  0.1987431354745856  & 0.5220218429952182  & -0.1926766886112917 \\
17&  0.2062478088988484  & 0.5304936855855431  & -0.1868475754585161 \\
18&  0.2133668331494581  & 0.5384793988419066  & -0.1812265016307355 \\
19&  0.2201380038925787  & 0.5460318047790646  & -0.1758031806343642 \\
20&  0.2265938267004754  & 0.5531955499144914  & -0.1705672082184823 \\
21&  0.2327624550322364  & 0.5600087116401930  & -0.1655083693982648 \\
22&  0.2386684311658198  & 0.5665040280450621  & -0.1606168213494539 \\
23&  0.2443332770354413  & 0.5727098525006191  & -0.1558831969805698 \\
24&  0.2497759696349747  & 0.5786509038428871  & -0.1512986565129531 \\
25&  0.2550133267993803  & 0.5843488625662739  & -0.1468549050440217 \\
26&  0.2600603227656853  & 0.5898228494962823  & -0.1425441882890768 \\
27&  0.2649303482326239  & 0.5950898137064368  & -0.1383592748238773 \\
28&  0.2696354261875951  & 0.6001648495876492  & -0.1342934304660328 \\
29&  0.2741863922040979  & 0.6050614580595203  & -0.1303403885609065 \\
30&  0.2785930459882241  & 0.6097917633368525  & -0.1264943186395051 \\
31&  0.2828642794964034  & 0.6143666940320125  & -0.1227497950252100 \\
32&  0.2870081858351735  & 0.6187961354134090  & -0.1191017663623826 \\
33&  0.2910321522986861  & 0.6230890581649763  & -0.1155455266354859 \\
34&  0.2949429402366960  & 0.6272536278701680  & -0.1120766879801252 \\
35&  0.2987467539279429  & 0.6312972985837565  & -0.1086911554135926 \\
36&  0.3024493002264770  & 0.6352268931891717  & -0.1053851035018207 \\
37&  0.3060558404258739  & 0.6390486727199947  & -0.1021549549113093 \\
38&  0.3095712355291510  & 0.6427683964162490  & -0.0989973607544213 \\
39&  0.3129999859059953  & 0.6463913739632261  & -0.0959091826148500 \\
40&  0.3163462661525911  & 0.6499225111032890  & -0.0928874761305482 \\
41&  0.3196139558344332  & 0.6533663496047807  & -0.0899294760096131 \\
42&  0.3228066666825220  & 0.6567271024057376  & -0.0870325823576189 \\
43&  0.3259277667231945  & 0.6600086846150990  & -0.0841943482007244 \\
44&  0.3289804017476282  & 0.6632147409439818  & -0.0814124680962679 \\
45&  0.3319675144656546  & 0.6663486700493243  & -0.0786847677306482 \\
46&  0.3348918616375112  & 0.6694136461978401  & -0.0760091944125418 \\
47&  0.3377560294345988  & 0.6724126385966856  & -0.0733838083775785 \\
48&  0.3405624472446620  & 0.6753484286860957  & -0.0708067748282852 \\
       \hline \hline
     \end{tabular}
    \end{center}
    \caption{1-loop coefficient $m^{(0)}_1(L)$ for improved actions.} 
    \label{fig:m1}
   \end{table}

We evaluated the 1-loop correction $m^{(0)}_1(L)$
numerically 
for the Iwasaki action ($c_1 = -0.331$, $c_2 = c_3 = 0$)
\cite{Iwasaki},
the L\"uscher-Weisz (LW) action 
($c_1 = -1/12$, $c_2 = c_3 = 0$)
\cite{LW}
and the DBW2 action
($c_1 = -1.40686$, $c_2 = c_3 = 0$)
\cite{DBW2} in the range $L = 6,\cdots,48$.
The results are shown
in Table \ref{fig:m1}.
The computations have been performed by using FORTRAN with 
the extended precision.
As a check of our calculation, we have confirmed
an independence of the gauge fixing parameter
and expected symmetries before reducing
the amount of the calculation \cite{su22loop}.
We have also checked that
our code at $c_1 = 0$ reproduces the known result of the Wilson plaquette 
action \cite{bode}.
Furthermore, two codes written independently  by the two authors
have produced identical results up to about 30 digits in the range 
$L = 6,\cdots,32$.
Beyond this range, we have used the faster code only.

 \section{Analysis and results}
  \label{sec:analysis}

In this section we extract the order $1/L$ term from the 1-loop
correction $m_1^{(0)}(L)$ to determine the O($a$) boundary improvement 
coefficient $c_{\rm{t}}^{(1)}$.
According to the Symanzik's analysis of the cutoff dependence
of Feynman diagrams on the lattice,
one expects that 1-loop coefficient has
an asymptotic expansion
    \begin{equation}
      m^{(0)}_1(L)
      \stackrel{L \rightarrow \infty}{\sim}
      \sum^{\infty}_{n=0}
      (A_n + B_n \ln L)/L^n.
     \label{eqn:m1zenkin}
    \end{equation} 
Using the blocking method of \cite{blocking},
we extracted the first few coefficients 
$A_0$, $B_0$, $A_1$, $B_1$ and estimated their errors.

Some of these coefficients are known or related to other quantities:
For example, $A_0$'s of two different actions are related to the ratio of 
$\Lambda$-parameters of two actions.
If the ultra-violet divergences in the SF is removed by
the standard renormalization of the coupling constant,
$B_0 = 2 b_0$, where $b_0 = 11/(4 \pi)^2$ is
the 1-loop coefficient of the $\beta$-function in the pure SU(3) gauge theory.
If the tree-level O($a$) improvement is implemented,
$B_1 = 0$ must hold.
Our main result comes from $A_1$: eq.(\ref{eqn:ctm1}) gives
$c_{\rm{t}}^{(1)}=A_1/2$.

We have first verified that our extraction of $B_0$ and $B_1$
is consistent with the above expectation.
We have confirmed $B_0 = 2 b_0$ up to 7 digits(Iwasaki), 9 digits(LW) or
4 digits(DBW2), while $B_1 < 10^{-4}$(Iwasaki), $< 10^{-7}$(LW) or $< 10^{-2}$
(DBW2).
Since our data give expected values of $B_0$ and $B_1$,
we fix $B_0 = 2 b_0$ and $B_1 = 0$ by hand,
in the blocking procedure to extract $A_0$ and $A_1$,
whose results for each action are shown in Table \ref{fig:A0A1}
where we have added the result of plaquette action
\cite{su3,bode}
for a later reference.

As a further check,
we extract $A_0$'s from the ratio of $\Lambda$-parameters
between two schemes $X$ and $Y$, which is given by
    \begin{equation}
      \frac{\Lambda_X}{\Lambda_Y}
      =
      e^{-\frac{c}{2 b_0}},
    \end{equation}
where
    \begin{equation}
      \bar{g}^2_Y(\mu)
      =
      \bar{g}^2_X(\mu)
      + c \bar{g}^4_X(\mu)
      + \cdots .
    \end{equation}
A purely numerical number $c$ here is given by
    \begin{equation}
    c=A_0^X-A_0^Y ,
    \end{equation}
where $A_0^X$ or $A_0^Y$ is the expected $A_0$ of the scheme $X$ or $Y$,
respectively.
We then find 
    \begin{equation}
      A^{\rm{imp}}_0
      =
      A^{\rm{plaq}}_0
      - 2 b_0 \ln 
      \daileft
        \frac{\Lambda_{\rm{imp}}}{\Lambda_{\rm{plaq}}}
      \dairight .
    \end{equation}     
Using $A^{\rm{plaq}}_0 = 0.36828215(13)$ \cite{su3,bode}
and the ratio of $\Lambda$-parameters 
\footnote{We take the value for 
DBW2 action from a private note \cite{saito}.}, 
    \begin{equation}
      \frac{\Lambda_{\rm{imp}}}{\Lambda_{\rm{plaq}}}
      =
      \left\{
        \begin{array}{ll}  
          59.05 \pm 1.0 
        & \mbox{for Iwasaki action \cite{Iwasakiyoshie}}
        \\
          5.29 \pm 0.01 
        & \mbox{for LW action \cite{Iwasakisakai}}
        \\
          1308 
        & \mbox{for DBW2 action \cite{saito}}
        \end{array}
      \right. ,
    \end{equation} 
we obtain the values of $A^{\rm{imp}}_0$ for each action,
which are shown in Table \ref{fig:A0A1} ($A^{\rm{exp}}_0$).
We have observed the consistency in $A_0$
between previous known results and our calculations.

With these confidences in our computation, 
we obtain the main result of our paper, the 1-loop O($a$) boundary
improvement coefficient eq.(\ref{eqn:ctm1}), which is given by
     \begin{equation}
       c^{(1)}_{\rm{t}}
       =
       c^{P(1)}_{\rm{t}}
       =
       A_1/2,
     \end{equation}
where $A_1$ is also found in Table \ref{fig:A0A1}.

Finally,
using our results $A_0$ in Table \ref{fig:A0A1},
we can estimate the ratio of $\Lambda$-parameters
between the improved action and the plaquette action
more accurately, 
    \begin{eqnarray}
      \frac{\Lambda_{\rm{imp}}}{\Lambda_{\rm{plaq}}}
      & = &
      \frac{\Lambda_{\rm{imp}}/\Lambda_{\rm{SF}}}
           {\Lambda_{\rm{plaq}}/\Lambda_{\rm{SF}}}
        =
      \exp 
        \left\{
          \frac{1}{2 b_0} [A_0^{\rm plaq} - A_0] 
        \right\}
      \nonumber \\
      & = &
      \left\{
        \begin{array}{ll}  
          61.2064(3) 
        & \mbox{for Iwasaki action}
        \\
          5.292104(5) 
        & \mbox{for LW action}
        \\
          1273.4(8) 
        & \mbox{for DBW2 action}
        \end{array}
      \right. .
    \end{eqnarray}

\section{Conclusion and discussions}
  \label{sec:ssf}

Combining our result of $c_{\rm{t}}^{(1)}$ for the improved gauge actions
and the previous result of $c_{\rm{t}}^{(1)}$ for the clover quark action \cite{sint},
we obtain
\begin{equation}
c_{\rm{t}}^{P(1)}=c_{\rm{t}}^{R(1)}/2 = A_1 /2 + n_f c_{\rm{t}}^{F(1)}
\end{equation}
for the $n_f$ flavors QCD, where $c_{\rm{t}}^{F(1)} = 0.0191410(1)$.

As a final remark, let us discuss the lattice artifact of the
step scaling function (SSF) \cite{O3ssf} for various gauge actions.
The SSF $\sigma(s,u)$ describes the 
evolution of a renormalized coupling under
finite rescaling factor $s$ (say $s=2$)
     \begin{equation}
       \sigma(s, u) 
       =
       \left.
         \bar{g}^2(sL) 
       \right|_{u = \bar{g}^2(L)},
     \end{equation}
and it has a perturbative expansion
     \begin{equation}
       \sigma(s,u)
       =
       u
     + 2 b_0 \ln (s) u^2
     + O(u^3).
     \end{equation}
This SSF $\sigma(s,u)$ in the continuum theory is obtained by the
continuum limit of the lattice SSF $\Sigma(s,u,1/L)$:
     \begin{equation}
       \sigma(s,u)
       =
       \lim_{1/L \rightarrow 0}
       \Sigma(s,u,1/L).
     \end{equation}
Therefore 
we can estimate the lattice artifact of the SSF in our perturbative
calculation.
We define the relative deviation $\delta (s,u,1/L)$ and
expand it as
     \begin{equation}
       \delta (2,u,1/L) 
       = 
         \frac{\Sigma(2,u,1/L) 
       - \sigma(2,u)}{\sigma(2,u)}
       = 
       \delta^{(k)}_1(2,1/L) u  
     + O(u^2),
     \end{equation} 
where we have set that $s = 2$ and
$\delta^{(k)}_1(2,1/L)$ is the 1-loop coefficient.
Here $k$ denotes the degree of the improvement for the boundary
coefficient: the tree (1-loop) value is used for $k=0$ ($k=1$).

   \begin{table} [p] 
\vspace{10mm}    
    \begin{center} 
     \begin{tabular}{|c|l|l|l|l|} 
       \hline \hline
         & plaquette action
         & Iwasaki action & LW action & DBW2 action 
       \\ \hline
           $A_0$            
         &\,\,\, $0.36828215(13)$ 
         & $-0.2049015(4)$ 
         &\,\,\, $0.136150567(6)$  
         & $-0.62776(8)$
       \\
           $A^{\rm{exp}}_0$ 
         & 
         & $-0.1999(24)$ 
         &\,\,\,\,\,$0.13621(26)$ 
         & $-0.6159$
       \\
           $A_1$
         & $-0.17800(10)$
         &\,\,\, $0.30360(26)$
         & $-0.005940(2)$   
         &\,\,\, $0.896(45)$
       \\
      \hline \hline
    \end{tabular}
    \end{center}
    \caption{The coefficients of asymptotic expansion $A_0$, $A_1$
             for various gauge actions. 
             The values for plaquette action are taken
from \cite{su3,bode}} 
    \label{fig:A0A1}
   \end{table}

   \begin{table} [p] 
    \begin{center} 
     \begin{tabular}{|c|c|c|c|c|c|c|c|c|} 
       \hline \hline
       & \multicolumn{2}{c|}{plaquette action} 
       & \multicolumn{2}{c|}{Iwasaki action} 
       & \multicolumn{2}{c|}{LW action} 
       & \multicolumn{2}{c|}{DBW2 action} 
       \\ \hline
       & & & & & & & & \\
       \raisebox{1.5ex}[0pt]{$L$}  
     & \raisebox{1.5ex}[0pt]{$\delta^{(0)}_1$} 
     & \raisebox{1.5ex}[0pt]{$\delta^{(1)}_1$} 
     & \raisebox{1.5ex}[0pt]{$\delta^{(0)}_1$} 
     & \raisebox{1.5ex}[0pt]{$\delta^{(1)}_1$} 
     & \raisebox{1.5ex}[0pt]{$\delta^{(0)}_1$} 
     & \raisebox{1.5ex}[0pt]{$\delta^{(1)}_1$} 
     & \raisebox{1.5ex}[0pt]{$\delta^{(0)}_1$} 
     & \raisebox{1.5ex}[0pt]{$\delta^{(1)}_1$} 
       \\ \hline
   6& 0.01089 & -0.00394 & -0.01922 & 0.00608 & 0.000911 & 0.000417 & -0.061 & 0.014 \\
   7& 0.01004 & -0.00268 & -0.01684 & 0.00484 & 0.000659 & 0.000236 & -0.050 & 0.014 \\
   8& 0.00918 & -0.00194 & -0.01499 & 0.00399 & 0.000527 & 0.000156 & -0.043 & 0.013 \\
   9& 0.00841 & -0.00148 & -0.01356 & 0.00330 & 0.000441 & 0.000111 & -0.038 & 0.011 \\
  10& 0.00773 & -0.00117 & -0.01241 & 0.00277 & 0.000379 & 0.000082 & -0.035 & 0.010 \\
  11& 0.00714 & -0.00095 & -0.01146 & 0.00235 & 0.000333 & 0.000063 & -0.032 & 0.009 \\
  12& 0.00663 & -0.00079 & -0.01064 & 0.00201 & 0.000296 & 0.000049 & -0.030 & 0.008 \\
  13& 0.00618 & -0.00066 & -0.00994 & 0.00174 & 0.000268 & 0.000039 & -0.028 & 0.007 \\
  14& 0.00579 & -0.00057 & -0.00932 & 0.00152 & 0.000244 & 0.000032 & -0.026 & 0.006 \\
  15& 0.00544 & -0.00049 & -0.00878 & 0.00134 & 0.000224 & 0.000026 & -0.024 & 0.006 \\
  16& 0.00513 & -0.00043 & -0.00830 & 0.00119 & 0.000207 & 0.000022 & -0.023 & 0.005 \\
  17& 0.00486 & -0.00038 & -0.00787 & 0.00106 & 0.000193 & 0.000018 & -0.022 & 0.005 \\
  18& 0.00461 & -0.00034 & -0.00748 & 0.00095 & 0.000181 & 0.000016 & -0.021 & 0.004 \\
  19& 0.00438 & -0.00030 & -0.00713 & 0.00086 & 0.000170 & 0.000013 & -0.020 & 0.004 \\
  20& 0.00418 & -0.00027 & -0.00681 & 0.00078 & 0.000160 & 0.000012 & -0.019 & 0.004 \\
  21& 0.00399 & -0.00025 & -0.00652 & 0.00071 & 0.000151 & 0.000010 & -0.018 & 0.003 \\
  22& 0.00382 & -0.00022 & -0.00625 & 0.00065 & 0.000144 & 0.000009 & -0.017 & 0.003 \\
  23& 0.00367 & -0.00020 & -0.00601 & 0.00059 & 0.000137 & 0.000008 & -0.017 & 0.003 \\
  24& 0.00352 & -0.00019 & -0.00578 & 0.00054 & 0.000131 & 0.000007 & -0.016 & 0.003 \\
     \hline \hline
     \end{tabular}
    \end{center}
    \caption{The deviations for various gauge actions.} 
    \label{fig:deviationIwasaki}
   \end{table}

   \begin{figure} [p]
     \begin{center}
      \resizebox{120mm}{!}{\includegraphics{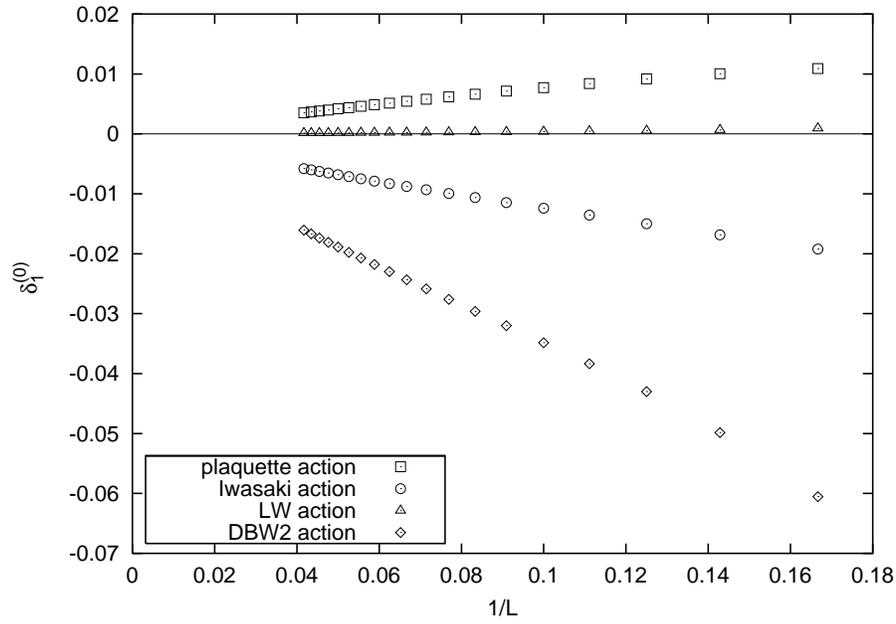}}
      \caption{The relative deviations of the lattice SSF 
               from the continuum 
               one at 1-loop for various gauge actions 
               with the tree-level O($a$) improved 
               boundary term.
One can find that $\delta^{(0)}_1$ for various gauge actions
vanish roughly linearly in 
$1/L$.}
      \label{fig:treelevelimpdeviation}
     \end{center}
    \end{figure}

    \begin{figure} [p]
      \begin{center}
       \resizebox{120mm}{!}{\includegraphics{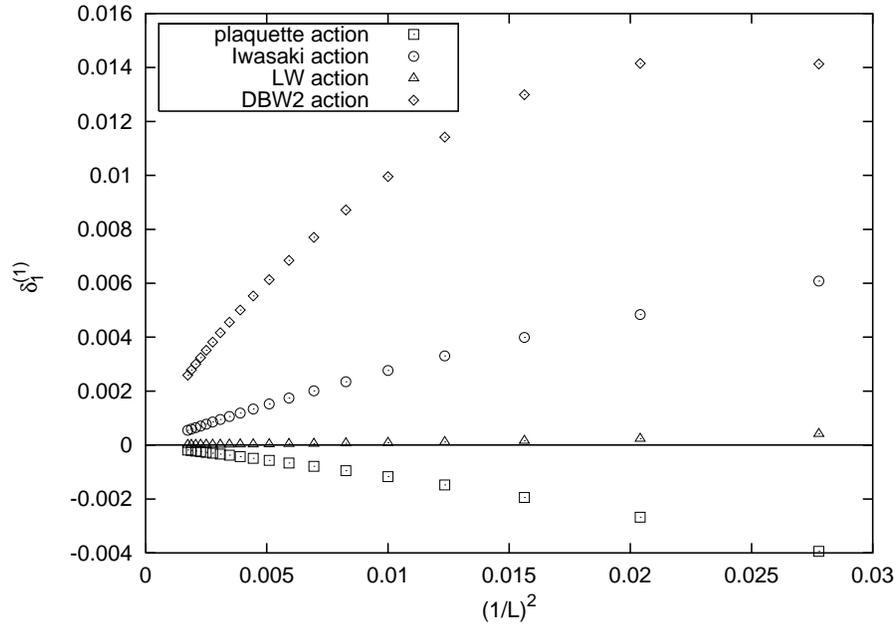}}
       \caption{The same quantities as in 
Fig.~\protect\ref{fig:treelevelimpdeviation}, but with the 1-loop level 
O($a$) improved boundary term.
One can find that $\delta^{(1)}_1$ for various gauge actions
vanish roughly quadratically in 
$1/L$.
}
       \label{fig:oneloopimpdeviation}
      \end{center}
    \end{figure}

The manifest form of $\delta^{(k)}_1(2,1/L)$
is given by
     \begin{equation}
       \delta^{(k)}_1(2,1/L)
       = 
       m_1^{(k)}(2L) 
     - m_1^{(k)}(L) 
     - 2 b_0 \ln(2),
      \label{eqn:deviation1}
     \end{equation}
and the results, including data of the plaquette action \cite{su3,bode}
for comparison\footnote{We have added data for the plaquette action
in the range $L=17,\cdots,24$},
are given in Table \ref{fig:deviationIwasaki}
for each gauge action.
Figure \ref{fig:treelevelimpdeviation}
(Figure \ref{fig:oneloopimpdeviation})
shows that 
the 1-loop deviations
with the tree-level (1-loop level) O($a$) improved boundary coefficient
vanish roughly linearly(quadratically) in 
$1/L$  as expected.
As is evident from Figure 1 and Figure 2,
at 1-loop level, 
the lattice artifact for the
renormalization group improved action (Iwasaki or DBW2)
is comparable to or larger than that for the plaquette action,
while the LW action is the least affected by the lattice cutoff.
However, one can not conclude that
the LW action is the best choice for numerical simulations,
where the lattice artifacts of higher orders in $u$ or $a$
may not be negligible.

 \section*{Acknowledgements}
S.T. would like to thank Dr. Saito for 
informative correspondence.

\newpage

\appendix


\renewcommand{\arraystretch}{1.5}

\section{Inverse propagator}
 \label{sec:inversepropagator}
Here, we give the explicit form of
the inverse propagator.
We choose the lattice unit (i.e. $a=1$) and set $T=L$.

    \begin{equation}
      K 
      = 
      c_{0} K^{(0)} 
    + c_{1} K^{(1)} 
    + \lam K^{(\rm{gf})}. 
    \end{equation}	

  \subsection*{Plaquette}
    \begin{eqnarray} 
      K^{(0)a}_{00}\pxy 
      & = & 
      R^a \delxy \mbox{\bf{s}}^a\px \cdot \mbox{\bf{s}}^a\pxpo, 
      \\
      &   & 
      \nonumber \\ 
      K^{(0)a}_{k0}\pxy 
      & = & 
      i R^a 
      \daileft 
        \delxy s^a_k\pxpo 
      - \delxmoy s^a_k\py 
      \dairight ,
      \\
      &   & 
      \nonumber \\
      K^{(0)a}_{0k}\pxy 
      & = & 
      - K^{(0)a}_{k0}\pyx, 
      \\
      &   & 
      \nonumber \\
      K^{(0)a}_{kl}\pxy 
      & = & 
      \delxy 
      \daileft 
        \delta_{kl} \mbox{\bf{s}}^a\px^2 
      - s^a_k\px s^a_l\px 
      \dairight 
      \nonumber \\
      &   & 
    + \delta_{kl} 
      \daileft 
        2 C^a \delxy 
      - R^a(\delxpoy + \delxmoy) 
      \dairight.
    \end{eqnarray}


  \subsection*{Rectangle}
       $w_{dbc}=3/2$ for the choice B \cite{aokiweisz}.
    \begin{eqnarray}   
      K^{(1)a}_{00}\pxy_{0kk} 
      & = & 
      \delxy 
      \daileft 
        1 + (w_{dbc} - 1) (\delxz + \delxTmo) 
      \dairight 
      \nonumber \\
      &   & 
      \times 4 R^a_2 \sum_m \sin(\phix + p_m) \sin(\phixpo + p_m), 
      \\
      &   & 
      \nonumber \\
      K^{(1)a}_{00}\pxy_{00k} 
      & = & 
      R^a_2 \sum_m 
      \chuuleft 
        \delxy  
        \daileft 
          ( 1 - \delxTmo ) s^a_m\px s^a_m\pxpt 
      \nonumber \\
      &   & 
      \mbox{\hspace{10mm}} 
      + ( 1 - \delxz )   s^a_m\pxmo s^a_m\pxpo 
        \dairight 
      \nonumber \\
      &   & 
      \mbox{\hspace{8mm}} 
      + \delxmoy s^a_m\pxmo s^a_m\pxpo 
      \nonumber \\ 
      &   & 
      \mbox{\hspace{8mm}} 
      + \delxpoy s^a_m\pymo s^a_m\pypo 
      \chuuright,
      \\
      &   & 
      \nonumber \\
      K^{(1)a}_{k0}\pxy_{0kk} 
      & = & 
      i R^a_2 2 c^a_k\px 
      \daileft 
        \delxy \{1 + (w_{dbc} - 1) \delxTmo \} \sin(\phixpo + p_k) 
      \nonumber \\
      &   & 
      \mbox{\hspace{17mm}} 
        - \delxmoy \{1 + (w_{dbc} - 1) \delxo \} \sin(\phiy + p_k) 
      \dairight, 
      \\
      &   & 
      \nonumber \\
      K^{(1)a}_{k0}\pxy_{00k} 
      & = & 
      i R^a_2 
      \chuuleft 
        s^a_k\pxpt [ ( 1 - \delxTmo )\delxy + \delxpoy ] 
      \nonumber \\ 
      &   & 
      \mbox{\hspace{3.5mm}} 
      - s^a_k\pxmt [ ( 1 - \delxo )\delxmoy + \delxmty ] 
      \chuuright, 
      \\ 
      &   & 
      \nonumber \\
      K^{(1)a}_{0k}\pxy 
      & = & 
    - K^{(1)a}_{k0}\pyx, 
      \\
      &   & 
      \nonumber \\
      K^{(1)a}_{kl}\pxy_{0kk} 
      & = & 
      \delta_{kl} c^a_k\px c^a_k\py 
      \daileft 
        2 C^a_2 \delxy - R^a_2 ( \delxpoy + \delxmoy ) 
      \dairight 
      \nonumber \\
      &   & 
      + ( w_{dbc} - 1 ) \delta_{kl} \delxy c^a_k\px 
      \daileft 
        \delxo ( C^a_2 c^a_k\px - i S^a_2 s^a_k\px) 
      \nonumber \\
      &   & 
      \mbox{\hspace{33mm}} 
      + \delxTmo ( C^a_2 c^a_k\px + i S^a_2 s^a_k\px) 
      \dairight,
      \\
      &   & 
      \nonumber \\
      K^{(1)a}_{kl}\pxy_{\rm{others}} 
      & = & 
      \delxy 
      \daileft 
        \delta_{kl} \sum_m s^a_m\px^2 ( c^a_k\px^2 + c^a_m\px^2 ) 
      \nonumber \\
      &   & 
      \mbox{\hspace{5mm}} 
      - s^a_k\px s^a_l\px ( c^a_k\px^2 + c^a_l\px^2 ) 
      \dairight 
      \nonumber \\
      &   & 
      \mbox{\hspace{-3mm}} 
    + \delta_{kl} 
      \daileft ( 2 - \delxo - \delxTmo ) C^a_2 \delxy 
      - R^a_2 ( \delxpty + \delxmty ) 
      \dairight.
    \end{eqnarray}

  \subsection*{Gauge fixing term}
    \begin{eqnarray}
      K^{({\rm{gf}})a}_{00}\pxy 
      & = & 
      2\delxy - \delxpoy - \delxmoy 
      \nonumber \\
      &   & 
      \nonumber \\
      &    & 
      - \delxy 
        [ \delxz ( 1 
        - \chi_a \delta_{\mbox{\scriptsize{\bf{p,0}}}} ) 
        + \delxTmo ], 
      \\
      &   & 
      \nonumber \\
      K^{({\rm{gf}})a}_{k0}\pxy 
      & = & 
      - i s^a_k\px [ \delxy - \delxmoy ], 
      \\
      &   & 
      \nonumber \\
      K^{({\rm{gf}})a}_{0k}\pxy 
      & = & 
    - K^{({\rm{gf}})a}_{k0}\pyx, 
      \\
      &   & 
      \nonumber \\
      K^{({\rm{gf}})a}_{kl}\pxy 
      & = & 
      \delxy s^a_k\px s^a_l\px.
    \end{eqnarray}

  \subsection*{Coefficients}
    \begin{equation}
      s^a_k\px 
      =
      2 \sin[ (p_k + \phix )/2 ],
    \end{equation}
    \begin{equation}
      c^a_k\px 
      = 
      2 \cos[ (p_k + \phix )/2 ],             
    \end{equation}
    \begin{equation}
      \phix 
      = 
    - \phi_{\bar{a}}(x_0),
    \end{equation}
    \begin{equation}
      C^a 
      = 
      C^{\bar{a}}, 
      \mbox{\hspace{3mm}} 
      S^a 
      = 
    - S^{\bar{a}},                               
      \mbox{\hspace{3mm}} 
      R^a 
      =
      R^{\bar{a}}, 
    \end{equation}
    \begin{equation}
      C^a_2 
      = 
      C^{\bar{a}}_2, 
      \mbox{\hspace{3mm}} 
      S^a_2 
      = 
    - S^{\bar{a}}_2,                               
      \mbox{\hspace{3mm}} 
      R^a_2 
      =
      R^{\bar{a}}_2, 
    \end{equation}
    \begin{equation}
      \chi_a 
      =
      \chi_{\bar{a}} 
      =
      (0,0,1,0,0,0,0,1),
    \end{equation}
    \begin{equation}
      \gamma 
      =
      \frac{1}{L^2} ( \eta + \frac{\pi}{3} ),   
    \end{equation}
where $\bar{1} = 2$, $\bar{4} = 5$, $\bar{6} = 7$, 
and vice verse. For diagonal part, $\bar{3} = 3$, $\bar{8} = 8$.


  \subsection*{Lists of coefficients}

   \begin{table}[h] 
    \begin{center} 
     \begin{tabular}{ccc} \hline \hline
       $a$ & $C^a$                                         
           & $S^a$                                         
           \\ \hline
       1,4 & $\frac{1}{2}(\cos 2 \gamma + \cos \gamma)$    
           & $-i \frac{1}{2}(\sin 2 \gamma + \sin \gamma)$ 
           \\ 
       3,6 & $\cos \gamma$                                 
           & $0$                                           
           \\         
       8   & $\frac{1}{3}(2 \cos 2 \gamma + \cos \gamma)$  
           & $0$                                           
           \\ \hline \hline
     \end{tabular}
    \end{center}
    \caption{$C^a$ and $S^a$ for SU(3).} 
   \end{table}

   \begin{table}[h] 
    \begin{center} 
      \begin{tabular}{ccc} \hline \hline
        $a$ & $\phix$                                        
            & $R^a$                                                 
            \\ \hline
        1   & $-3 \gamma x_0 
              + \frac{1}{L} 
                ( \eta [\frac{3}{2} - \nu] - \frac{\pi}{3})$   
            & $\cos \frac{\gamma}{2}$                             
            \\ 
        4   & $-3 \gamma x_0 
              + \frac{1}{L} 
                ( \eta [\frac{3}{2} + \nu] - \frac{2\pi}{3})$  
            & $\cos \frac{\gamma}{2}$   
            \\         
        3   & $0$                  
            & $\cos \gamma$                                       
            \\
        6   & $\frac{1}{L}(2 \eta \nu - \frac{\pi}{3})$    
            & $\cos \gamma$                                       
            \\ 
        8   & $0$                 
            & $\frac{1}{3}(2 \cos 2 \gamma + \cos \gamma)$     
            \\ \hline \hline
      \end{tabular}
     \end{center}
    \caption{$\phix$ and $R^a$ for SU(3).} 
   \end{table}

   \begin{table}[h] 
    \begin{center} 
     \begin{tabular}{cccc} \hline \hline
        $a$ & $R^a_2$                                         
            & $C^a_2$                 
            & $S^a_2$ 
            \\ \hline
        1,4 & $\cos \gamma$ 
            & $R^a_2 \cos 3 \gamma$   
            & $-i R^a_2 \sin 3 \gamma$  
            \\ 
        3,6 & $\cos 2 \gamma$                                 
            & $R^a_2$                 
            & $0$                                            
            \\         
        8   & $\frac{1}{3}(2 \cos 4 \gamma + \cos 2 \gamma)$  
            & $R^a_2$                 
            & $0$                                            
            \\ \hline \hline
      \end{tabular}
     \end{center}
    \caption{$R^a_2$, $C^a_2$ and $S^a_2$ for SU(3).} 
   \end{table}

\end{document}